\documentclass[conference]{IEEEtran}
\usepackage{graphicx}
\usepackage{url}
\begin{document}

\title{A Bibliometric Study of Asia Pacific Software Engineering Conference from 2010 to 2015}

\author{\IEEEauthorblockN{Lov Kumar}
\IEEEauthorblockA{NIT Rourkela, India\\
lovkumar505@gmail.com}
\and
\IEEEauthorblockN{Saikrishna Sripada}
\IEEEauthorblockA{IIIT Hyderabad, India\\
saikrishna.sripada@research.iiit.ac.in}
\and
\IEEEauthorblockN{Ashish Sureka}
\IEEEauthorblockA{ABB, India\\
ashish.sureka@in.abb.com}}
\maketitle
\begin{abstract}
The Asia-Pacific Software Engineering Conference (APSEC) is a reputed and a long-running conference which has successfully completed more than two decades as of year $2015$. We conduct a bibliometric and scientific publication mining based study to how the conference has evolved over the recent past $6$ years (year $2010$ to $2015$). Our objective is to perform in-depth examination of the state of APSEC so that the APSEC community can identify strengths, areas of improvements and future directions for the conference. Our empirical analysis is based on various perspectives such as: paper submission acceptance rate trends, conference location, scholarly productivity and contributions from various countries, analysis of keynotes, workshops, conference organizers and sponsors, tutorials, identification of prolific authors, computation of citation impact of papers and contributing authors, internal and external collaboration, university and industry participation and collaboration, measurement of gender imbalance, topical analysis, yearly author churn and program committee characteristics. 
\end{abstract}
\begin{IEEEkeywords}
Asia-Pacific Software Engineering Conference, Bibliometric Analysis, Mining Scientific Papers, Software Engineering Research Reflection
\end{IEEEkeywords}
\IEEEpeerreviewmaketitle

\section{Research Motivation and Aim}
The Asia-Pacific Software Engineering Conference (APSEC) is an annual conference started in the year $1994$ in Tokyo (Japan) with the aim of bringing software engineering researchers and practitioners from both university and industry around the world to exchange research results and ideas. APSEC $1994$ (Tokyo, Japan) was the first edition of the conference and the annual event completes $23$ editions in $2016$ (Hamilton, New Zealand). More than two decades of APSEC has been successful in providing a forum to share cutting-edge advancements in the field of software engineering and has become one of the important academic events for researchers and scholars (particularly in the Asia-Pacific region) working in the area of software engineering. The number of national and international delegates participating in APSEC series of conferences every year, citation impact of the published scientific papers (research and industry track), quality of peer-reviews, number of papers submitted to the various conference tracks, diversity and quality of co-located events such as workshops and tutorials, keynotes and invited talks, professional reputation of the organizing and program committee members and support from industry as well as government agencies as sponsors are clear evidences indicating that APSEC has maintained its reputation and high quality. We believe that a reflection of the recent past (year $2010$ to $2015$) of APSEC is important for the APSEC community for learning and further improving the impact and quality of the conference.

The bibliometric and scientific publication mining based study presented in this paper is motivated by the need to conduct a quantitative and objective evaluation and assessment of past $6$ years of APSEC. Our research motivation is to investigate answers to questions such as: how the conference has evolved over the past $6$ years and what is the current status, what is the quality of the conference based on several key performance indicators, what improvements can be made and to what extent APSEC is meeting its desired objectives. Our research aim is to systematically and scientifically explore and examine the state of APSEC across various aspects of the conference. To the best of our knowledge, the study presented in this paper is the first in-depth examination of the state of APSEC which we believe is important for the APSEC community to understand its development, evolution and identify future directions.
\section{Related Work and Research Contributions}
In this Section, we present related work and the novel contributions of the study presented in this paper in context to existing work. Tripathi et al analyze research papers published in MSR (Mining Software Repositories) series of conferences from $2010$ to $2014$ (a period of $5$ years) \cite{tripathi2015}. Sharma et al. conduct a bibliometric study consisting of mining $551$ papers in Requirements Engineering (RE) series of conference of \cite{sharma2016}. Sharma et al. analyze $11$ years of RE papers published from the year $2005$ to $2015$. They study several aspects such as: authorship numbers and scholarly productivity of various countries or regions, interdisciplinarity, topic modeling and categorization, collaboration (university and industry, internal and external) and public and proprietary dataset \cite{sharma2016}. Agarwal et al. conduct a research study on gender imbalance and low participation of women in Computer Science Research (CSR) \cite{swati2016}. They conduct several empirical and statistical analysis consisting of mining thousands of bibliometric entries in DBLP\footnote{\url{ http://dblp.uni-trier.de/}} bibliography data \cite{swati2016}. Their findings reveal that in the broad field of Computer Science, there is a gender balance wherein only $21\%$ of the authors are female and $79\%$ are male authors. They study several aspects of the conference like national and international collaboration, university and industry collaboration, research type of the study (validation, evaluation, solution proposal, philosophical, opinion and experience), contribution (such as tooling or metric) and technology area \cite{barn2016}.
Barn et al. present a systematic mapping study to all the papers published in Indian Software Engineering Conference (ISEC) series since its beginning \cite{barn2016}.  Freitas et al. perform a bibliometric analysis of $740$ scientific papers published in Search Based Software Engineering (SBSE) series of conference from the year $2001$ to $2010$ \cite{freitas2011}. Bergamaschi present a quantitative analysis of WWW, Hypertext (HT) and JCDL which are three of the seven ACM SIGWEB sponsored conferences \cite{bergamaschi2012}. Bartneck present a historical overview of the long-running and prestigious conference on Human Robot Interaction (HRI) \cite{bartneck2010}. Bartneck et al. present a scientometric analysis of the CHI (one of the top-most conference on Human Computer Interaction) proceedings \cite{bartneck2009}. Agarwal et al. present a bibliometric analysis of the scientific publications and corresponding ACM metadata (also published as a research output) of seven conferences sponsored by ACM SIGWEB \cite{agarwalsigweb2016}.
\\\\
\textbf{Research Contributions:} In context to existing work, the study presented in this paper makes several novel research contributions. The work presented in this paper is the first in-depth and focused study on analyzing Asia-Pacific Software Engineering Conference (APSEC) across various perspectives: paper acceptance rates, conference location, scholarly output of various countries, keynotes, workshops, conference organizers and sponsors, tutorials, prolific authors, citation impact, internal and external collaboration, university and industry participation and collaboration, gender imbalance, topical analysis, yearly author churn and program committee characteristics.
\\\\
\textbf{Dataset Contributions:} We make our dataset publicly available on figshare\footnote{\url{ https://figshare.com/}} which is a web platform for storing, sharing and discovering research \cite{sureka2016}. We believe that sharing our dataset will further facilitate research on APSEC bibliometric analysis and can be used to explore new research problems and hypothesis. Extracting data from a large number of papers required manual effort as well as writing scripts and web believe that the research community can benefit from our shared data.  
\\\\
\textbf{Extended Version of Previous Paper:} The study presented in this paper is an extended version of the short paper accepted in APSEC 2016 by the same authors \cite{lov2016}. Due to the four page limit of the APSEC 2016 paper, several aspects are not covered which are described in this paper. The objective of this paper is to provide a complete and detailed analysis of our work on APSEC bibliometric analysis (year $2010$ to $2015$) through arXiv open access\footnote{\url{https://arxiv.org/}}. 
\begin{figure}[th]
\centering
\includegraphics[width=0.47\textwidth]{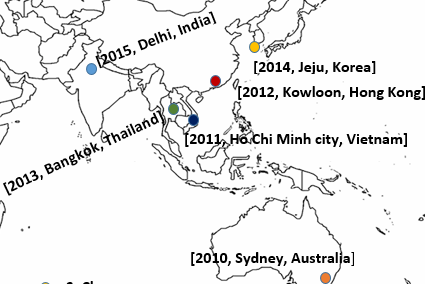}
\caption{Map of Asia-Pacific Region displaying the Conference Location, Year and General Chairs for APSEC $2010$ to APSEC $2015$}
\label{apsecmap}
\end{figure}
\begin{table}[ht]
\caption{Number of Papers Submitted (SUB), Research Regular (REG), Research Short ot Poster (SHR), Industry Track (IND), Emerging Research (ERA), Accepted (ACC) and Acceptance Rate (AR) APSEC from $2010$-$2015$}
\label{accsub}
\centering
\renewcommand{\arraystretch}{1.1}
\resizebox{8.50cm}{!}{
\begin{tabular}{|c|c|c|c|c|c|c|c|} 
\hline
\textbf{Year} & \textbf{REG} & \textbf{SHR} & \textbf{IND} & \textbf{ERA} & \textbf{ACC} & \textbf{SUB} & \textbf{AR}  \\ \hline
\textbf{2010} & 47 & 0 & $17^{*}$ & 0 & 47 & $151$ & $47/151=31.12\%^{*}$ \\ \hline
\textbf{2011} & 49 & 0 & 0 & 0 & 49 & $178$ & 49/178=27.52\% \\ \hline
\textbf{2012} & 51 & 48 & \multicolumn{2}{|c|}{$23^{*}$} & 99 & $200$ & $99/200=49.5 \%^{*}$ \\ \hline
\textbf{2013} & 61 & 14 & $12^{*}$ & - & 75 & $167$ & $75/167=44.91\%^{*}$ \\ \hline
\textbf{2014} & 58 & 4 & 7 & - & 69 & $226$ & 69/226=30.53 \% \\ \hline
\textbf{2015} & 42 & - & 9 & - & 51 & $182$ & 51/182=28.02\% \\ \hline
\end{tabular}}
\end{table}
\section{Location, Acceptance Rate and Organizers}
Figure \ref{apsecmap} shows the city, country and year for APSEC $2010$ to $2015$. The map in Figure \ref{apsecmap} shows that the conference is being organized in different countries in the Asia-Pacific region and hence there is a geographical balance in-terms of the conference location. We observe that from the year $2010$ to $2015$, the conference is generally being organized in one of the largest cities of the hosting country except in the year $2014$ (Jeju Island, Korea). We download all the APSEC $2010$ to $2015$ papers from the IEEE digital library for our analysis. We compute the acceptance rate and number of papers submitted each year based on the papers available in the IEEE digital library and the information provided by the program chairs in the front matter or message for the program chair(s) section of the proceedings. We observe several inconsistencies and incompleteness of the data in message from the program chairs.  For example, for the year $2013$, the number of submission are provided only for the research track and not for other tracks. For the year 2010 there are no industry track in the proceedings but the website shows an industry track. For the year $2012$, we compute acceptance rate based on the $99$ research track paper ($51$ full papers and $48$ short papers) as the submission data of ERA and IND track is not given. 

Table \ref{accsub} shows the number of papers accepted and submitted across various tracks and the acceptance rate or selectivity of the conference. Table \ref{accsub} reveals that the acceptance rate varies from a minimum of $27.52\%$ to a maximum of $49.5\%$. Table \ref{accsub} shows that the number of papers submitted to APSEC falls in the range of $150$ to $230$. We observe that several tracks like short papers, industry track and emerging research track are not conducted every year and are not regular features of APSEC. Table \ref{conferenceinfo} displays the Conference Location, General Chair(s), Program Chair(s) and Sponsors of APSEC from $2010$ to $2015$. Table \ref{conferenceinfo} shows that the conference leadership (general and program Chairs) have representation mostly from university and rarely from industry (except in the year $2015$ when both the general and program chairs were from industry). The list of sponsors mentioned in Table \ref{conferenceinfo} that APSEC receives sponsorship and support from all the sectors: university, industry and government. 
\begin{table*}[htpb]
\centering
\caption{Conference Location, General Chair(s), Program Chair(s) and Sponsors of APSEC from 2010 to 2015}
\label{conferenceinfo}
\renewcommand{\arraystretch}{1.1}
\resizebox{18.5cm}{!}{
\begin{tabular}{|c|l|p{4.5cm}|p{4.5cm}|p{5.5cm}|}
\hline
\textbf{Year} & \textbf{Location} & \textbf{General Chair} & \textbf{Program Chair} & \textbf{Sponsors} \\ \hline
2015 & New Delhi, India & Srinivas Padmanabhuni, Infosys, India; Ashish Sureka, ABB Corporate Research, India & Jing Sun University of Auckland, New Zealand; Y. Raghu Reddy IIIT Hyderabad, India & Infosys, TCS, ACM India, ISOFT \\ \hline
2014 & Jeju, Korea & Sungdeok (Steve) Cha, Korea University, Korea & Yann-Gaël Guéhéneuc, Polytechnique Montréal, Canada; Gihwon Kwon, Kyonggi University, Korea & Kisti, Nipa, Etri, LG Electronics, Samsung Electronic, Solution Link, SSDRC,  ESG, Suresoft \\ \hline
2013 & Bangkok, Thailand & Prabhas Chongstitvatana, Chulalongkorn University, Thailand; Pansak Siriruchatapong, NECTEC, Thailand & Pornsiri Muenchaisri, Chulalongkorn University, Thailand; Gregg Rothermel, University of Nebraska, USA & TCEB, Provincial Electricity Authority, Meteropolitan Electricity Authority \\ \hline
2012 & Kowloon, Hong Kong & T.H. Tse, The University of Hong Kong, China; S.C. Cheung, The Hong Kong University of Science and Technology, China & Karl Leung, The Hong Kong Institute of Vocational Education, China; Pornsiri Muenchaisri, Chulalongkorn University, Thailand & IEEE Hong Kong Section Computer Society Chapter, ACM Hong Kong Chapter, ACM Chapter Conference \\ \hline
2011 & Ho Chi Minh city, Vietnam & Bach Hung Khang, Institute of Information Technology, Vietnam; Duong Anh Duc, University of Science, VNU-HCM, Vietnam  & Tran Dan Thu, University of Science, VNU-HCM, Vietnam; Karl Leung, Hong Kong Institute of Vocational Education, China  & Department of Information and Communication, HCM City Information Sponsor, Information Technology Center - University of Science HCMU Financial Sponsor,TMA Solution - Leading Software Outsoursing Company in Vietnam Sponsor for Gifts of Industrial Section \\ \hline
2010 & Sydney, Australia & Paul Strooper, The University of Queensland, Australia; Ross Jeffery, National ICT Australia (NICTA), Australia  & Jun Han, Swinburne University of Technology, Australia; Tran Dan Thu, University of Science - HCMC VNU, Vietnam  & National ICT Australia (NICTA), IBM, Microsoft Research , Australian Safety Critical Systems Association, Swinburne University of Technology, The University of Queensland \\ \hline
\end{tabular}}
\end{table*}
\begin{table*}[ht]
\caption{List of Keynotes at APSEC $2010$ to APSEC $2015$}
\label{keynotes}
\centering
\renewcommand{\arraystretch}{1.1}
\resizebox{18.0cm}{!}{
\begin{tabular}{|c|p{10cm}|l|} 
\hline
\textbf{Year} & \textbf{Title} & \textbf{Speaker}  \\ \hline
 & The Data Deluge, How Software Engineering can Help & Judith Bishop, Microsoft (USA) \\
2010 & Changes in the Software Development Profession to Meet the Need for Innovation from Businesses and Government & Martin Nally, IBM (USA) \\ \hline
2011 & Future Software Engineering Opportunities and Challenges & Barry W. Boehm, University of Southern California, USA \\
& Software Process Definition & Leon J. Osterweil, University of Massachusetts, USA \\ \hline
 & Whither Software Architecture? & Jeff Kramer, Imperial College London, UK \\
2012& Whither Software Engineering Research? & David S. Rosenblum,  National University of Singapore, Singapore \\
 & Blended Program Analysis for Improving Reliability of Real-world Applications & Barbara G. Ryder, Virginia Tech, USA \\ \hline
 & Useful Software Engineering Research: Leading a Double-Agent Life & Lionel C. Briand, FTSC, University of Luxembourg  \\
2013& Practices of Software Engineering in Thailand & Dr. Kanchit  Malaivongs, Royal Institute of Thailand \\
& Achieving Success in Open Software Ecosystems:  The Role of Architectural Styles &  Richard Taylor, University of California, Irvine \\ \hline
 & Strategy Introduction of Embedded Software Competence for IoT era & Jeonghan Kim(Brian), Samsung Electronics, Korea \\
2014& Powers of Two: Cultures, Solitudes and Software Engineering   & Mike Hoye, Mozilla, Canada \\
& Architecting = Decision Making & Hans van Vliet, VU University Amsterdam \\ \hline
 & Evolving Critical Systems & Dr Minke Henchey, University of Limerick (UL), LERO \\
2015& Building an Open Identity Platform for India & Dr Pramod Varma, Chief Architect, UIDAI \\ 
& Trends in Automation and Control Systems – Needing Efficient Software Engineering & Mr Akilur Rahman, Head of ABB INCRC, India \\ \hline
\end{tabular}}
\end{table*}
\begin{table}[h]
\centering
\caption{List of Workshops at APSEC $2010$ to APSEC $2015$}
\label{workshop}
\renewcommand{\arraystretch}{1.1}
\resizebox{9cm}{!}{
\begin{tabular}{|c|c|p{6.9cm}|}
\hline
\textbf{Year} & \textbf{No.} & \textbf{Workshop}  \\ \hline
2010 & 1 & Cloud Computing  \\ 
& 2 & Software Engineering for Embedded Systems \\ \hline
2011 & 1 & International Workshop on Recent Progress in Software Engineering \\ \hline
2012 & 1 & International Workshop on Software Analysis, Testing and Applications \\ 
& 2 & International Workshop on Software Quality and Management \\ \hline
2013 & 1 & The 5th International Workshop on Empirical Software Engineering in Practice (IWESEP 2013) \\ 
& 2 & International Workshop on Quantitative Approaches to Software (QuASoQ 2013) \\ \hline
2014 & 1 & 2nd International Workshop on Quantitative Approaches to Software Quality (QuASoQ 2014) \\ 
& 2 & Software Engineering Education Workshop (SEEW) \\ \hline
& 1 & Quantitative Approaches to Software Quality (QuASoC) \\ 2015 & 2 & Alternate Workforces for Software Engineering (WAWSE) \\ 
& 3 & Case Method for Computing Education (CMCE) \\ \hline
\end{tabular}}
\end{table}	
\begin{table}[h]
\centering
\caption{List of Tutorials at APSEC $2010$ to APSEC $2015$}
\label{tutorials}
\renewcommand{\arraystretch}{1.1}
\resizebox{9cm}{!}{
\begin{tabular}{|c|c|p{7.2cm}|}
\hline
\textbf{Year} & \textbf{No.} & \textbf{Tutorial}  \\ \hline
2011 & 1 & Requirements Engineering Based on Requirements Engineering Body Of Knowledge (REBOK)  \\
& 2 & A Process Decision Framework: The Incremental Commitment Spiral Model  \\ \hline
& 1 & A Survey of Domain Engineering  \\
2012 & 2 & Requirements Engineering Based on REBOK (Requirements Engineering Body of Knowledge) and its Practical Guide  \\
& 3 & Product Line Requirements Reuse based on Variability Management  \\ \hline
& 1 & Requirements Engineering Based on REBOK (Requirements Engineering Body Of Knowledge and its Practice  \\
2013 & 2 & SAT and SMT their algorithm designs and applications  \\
& 3 & Model‐based Transition from Requirements to High‐level Software Design  \\
& 4 & Software Reuse based on Business Processes and Requirements  \\ \hline
& 1 & Metamorphic Testing  \\
2014 & 2 & Using Artificial Intelligence Techniques for Requirements Engineering Research  \\
& 3 & Embedded Software Design in the Hardware/Software Codesign Methodology  \\ \hline
2015 & 1 & Building Enterprise-grade Internet of Things Applications  \\
& 2 & Software Defined Storage Technology  \\ \hline
\end{tabular}}
\end{table}	
\section{Keynotes, Workshops and Tutorials}
Table \ref{keynotes} shows the list of keynotes at APSEC $2010$ to APSEC $2015$. We extract the keynote information from the extended abstract for keynote included as part of the conference proceedings. We observe a good mix of keynotes from industry (such as ABB, IBM, Microsoft and Samsung), government (such as UIDAI) and university. We notice a talk from non-profit corporations like Mozilla Foundation.  Table \ref{keynotes} shows that keynote speakers are from various parts of the world and are invited from countries outside the host country which shows conference diversity and inclusiveness.  Table \ref{workshop} shows the list of workshops at APSEC $2010$ to APSEC $2015$. We observe co-located workshops to APSEC on diverse topics such as cloud computing, embedded systems, software testing, software quality, empirical software engineering, software engineering education, alternate workforces and quantitative approaches to software quality. Normally, there are two co-located workshops to APSEC expect for the year $2011$ when there was only one co-located workshop and year $2015$ when there was $3$ co-located workshops. Table \ref{tutorials} shows the list of tutorials at APSEC $2010$ to APSEC $2015$. Table \ref{tutorials} reveals that there are at-least $2$ tutorials conducted with APSEC every-year and the tutorial covers diverse topics within software engineering.
\section{Citation Based Impact}
One of the major factors used to evaluate the quality of a conference is to measure the overall impact of the papers published or presented in the conference. The number of citations or references in other publications received by a scientific paper is an indicator of the impact of the paper. Google Scholar\footnote{url{http://scholar.google.com/}} is a popular and widely used search engine as well as a citation metrics service which keeps tracks of citations to articles indexed by it. We compute the number of citations for all the papers published in APSEC from the year $2010$ to $2015$. The total number of citations as of $5$ May $2016$ in year $2010$, $2011$, $2012$, $2013$, $2014$ and $2015$ are $642$, $340$, $386$, $179$, $53$ and $0$ respectively. The total number of citations from year $2010$ to $2016$ is $1600$. The h5-index for APSEC on $5$ May $2016$ is $13$. Google Scholar defines h5-index as "h5-index is the h-index for articles published in the last 5 complete years. It is the largest number h such that h articles published in 2010-2014 have at least h citations each". The h5-median for APSEC on $5$ May $2016$ is $18$. Google Scholar defines h5-median as "h5-median for a publication is the median number of citations for the articles that make up its h5-index". 

Table \ref{top10citedpaper} shows the Top $10$ most cited APSEC papers in our dataset. Table \ref{top10citedpaper} reveals that only one paper has received more than $50$ citations and $6$ out of $10$ have received more than $25$ citations. The minimum number of citations received by the Top $10$ papers is $20$. Table \ref{top10citedpaper} shows the first author name and country of affiliation of the first authors. The data in Table \ref{top10citedpaper} shows a wide diversity of scholarly output and contributions to APSEC from various countries. Garousi et al. \cite{garousi2016highly} conducted a study, comprised of five research questions, to identify and classify the top-100 highly-cited SE papers in terms of two metrics: total number of citations and average annual number of citations. According to their study International Conference on Software Engineering (ICSE), International Conference on Requirements Engineering (RE), Conference on Object-Oriented Programming Systems, Languages, and Applications (OOPSLA), IEEE/ACM/IFIP International Conference on Hardware/Software Codesign and System Synthesis (CODES+ISSS) and IEEE/ACM International Conference on Automated Software Engineering (ASE) are the top five conference venues for software engineering papers. APSEC does not appear in the list of conferences under the category of venues for top papers. The h5-index for APSEC on $5$ May $2016$ is $13$ whereas the h5-index\footnote{\url{http://2015.msrconf.org/MSR_Impact.pdf}} of several broad scoped software engineering conferences computed by the MSR committee on December 31, 2014 is: ICSE (57), FSE (38), ASE (30) and ESEM (24). 
\begin{table*}[ht]
\caption{Top $10$ Most Cited APSEC $2010$ to $2015$ Papers Based on Google Scholar Metrics (Citations Metrics Collected on 5 May 2016)}
\label{top10citedpaper}
\centering
\renewcommand{\arraystretch}{1.1}
	\resizebox{18.0cm}{!}{
 \begin{tabular}{|c|c|p{11.5cm}|l|l|c|} 
 \hline
 \textbf{Rank} & \textbf{Year} & \textbf{Paper Title} & \textbf{First Author} & \textbf{Country} & \textbf{Citations} \\ \hline
1 & $2010$ & The Qualitas Corpus: A Curated Collection of Java Code for Empirical Studies & Ewan Tempero & New Zealand & 173 \\ \hline
2 & $2010$ & Detecting Duplicate Bug Report Using Character N-Gram-Based Features & Ashish Sureka & India & 42 \\ \hline
3 & $2011$ & DREX: Developer Recommendation with K-Nearest-Neighbor Search and Expertise Ranking & Wenjin Wu & China
& 41 \\ \hline
4 & $2010$ & Evaluating Cloud Platform Architecture with the CARE Framework & Liang Zhao & Australia
 & 36 \\ \hline
5 & $2010$ & Approaching Non-functional Properties of Software Product Lines: Learning from Products & Julio Sincero & Germany
 & 35\\ \hline
6 & $2012$ & A Design Pattern to Build Executable DSMLs and Associated V and V Tools & Benoit Combemale & France
 & 27 \\ \hline
7 & $2010$ & Model-Based Methods for Linking Web Service Choreography and Orchestration & Jun Sun & Singapore
 & 24 \\ \hline
8 & $2010$ & Evaluating Mutation Testing Alternatives: A Collateral Experiment & Marinos Kintis & Greece
 & 22 \\ \hline
9 & $2013$ & Suggesting Extract Class Refactoring Opportunities by Measuring Strength of Method Interactions & Giuseppe Pappalardo & Italy
 & 21 \\ \hline
10 & $2010$ & Bridging the Gap between Fault Trees and UML State Machine Diagrams for Safety Analysis & HyeonJeong Kim & South Korea
 & 20 \\ \hline
\end{tabular}}
\end{table*}
\begin{table}[ht]
\caption{Descriptive Statistics for APSEC $2010$ to $2015$ Google Scholar Citations}
\label{citetable}
\centering
\begin{tabular}{|c|c|c|c|c|c|c|c|} 
\hline
\textbf{Year} & \textbf{Min.} & \textbf{Max.} & \textbf{Mean} & \textbf{Med} & \textbf{Skew} & \textbf{Kurt} & \textbf{Sum} \\ \hline
2010 & 0 & 173 & 13.66 & 9 & 5.37 & 33.84 & 642 \\ \hline
2011 & 0 & 41 & 6.94 & 6 & 2.90 & 14.77 & 340 \\ \hline
2012 & 0 & 27 & 3.16 & 2 & 2.87 & 16.72 & 386 \\ \hline
2013 & 0 & 21 & 2.06 & 1 & 4.10 & 24.86 & 179 \\ \hline
2014 & 0 & 6 & 0.77 & 0 & 1.89 & 6.97 & 53 \\ \hline
2015 & 0 & 0 & 0.00 & 0 & 0.00 & 0.00 & 0 \\ \hline
All & 0 & 173 & 4.28 & 2 & 11.92 & 186.71 & 1600 \\ \hline
\end{tabular}
\end{table}
Table \ref{citetable} displays the descriptive statistics for APSEC $2010$ to $2015$ Google Scholar Citations. The mid-point (median value) in-terms of the number of citations for the year $2010$ is $9$ and for the year $2011$ is $6$. Table \ref{citetable} reveals that half of the papers in APSEC $2013$ have citations equal to $1$ or less than $1$. We compute the skewness and kurtosis metrics to measure the symmetry (or lack of symmetry) and whether the citation data is heavy-tailed or light-tailed relative to a normal distribution. We observe that distribution for the years is positively skewed and hence the probability density function has a long tail to the right. We observe that the distribution is highly skewed for the year $2010$, $2011$, $2012$ and $2013$ and moderately skewed for the year $2014$. We observe that the probability density function for the year $2010$ and $2013$ has a higher kurtosis (more peaked at the center) than the kurtosis value for the year $2011$ and $2012$. Year $2014$ has a lower kurtosis value in comparison to all other years which means that the distribution for the year $2014$ is less peaked or more flat relative to other years. 

\section{Scholarly Output of Countries}
\begin{figure}[th]
\centering
\includegraphics[width=0.47\textwidth]{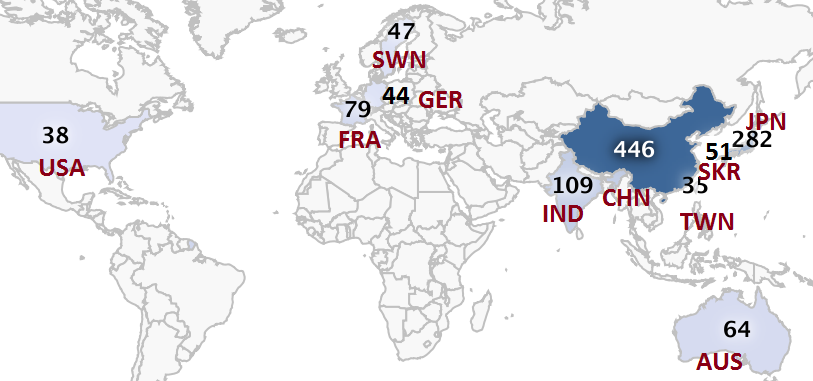}
\caption{Top $10$ Countries with the Highest Scholarly Output and Contributions}
\label{top10}
\end{figure}

We analyze the scholarly output of various countries contributing to the success of APSEC. We extract the country of affiliation for every co-author. If an author has published multiple papers in a particular year or across years then we award multiple points to the affiliated countries. Our analysis reveals that authors from $37$ different countries have contributed to APSEC from the year $2010$ to $2015$. The Top $3$ countries ($8.1\%$ of the countries) have a contribution of $57.68\%$ and the Top $5$ countries ($13.51\%$) have a contribution of $67.53\%$. The graph in Figure \ref{top10} displays the Top $10$ countries with the highest scholarly output and contributions. 

\begin{figure}[th]
\centering
\includegraphics[width=0.47\textwidth]{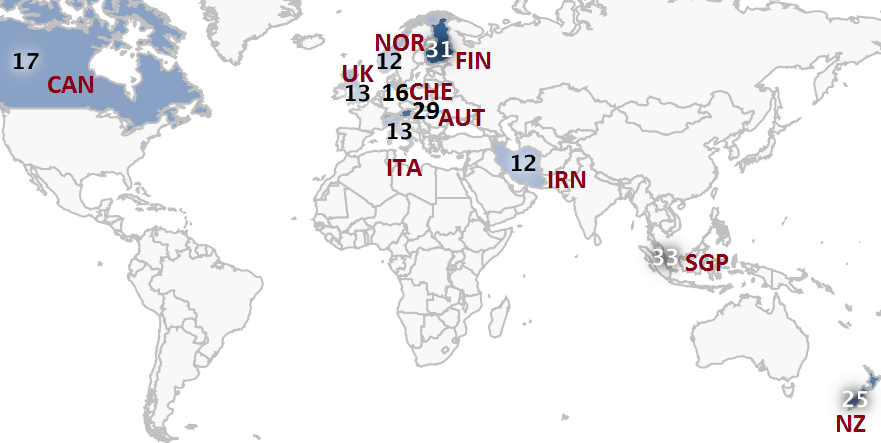}
\caption{Top $11$ to $20$ Countries with the Highest Scholarly Output and Contributions}
\label{top20}
\end{figure}

The Top 10 countries are: China (CHN, $446$) Japan (JPN, $282$), India (IND, $109$), France (FRA, $79$), Australia (AUS, $64$), Sweden (SWN, $47$), Germany (GER, $44$), South Korea (SKR, $51$), USA (USA, $38$) and Taiwan (TWN, $35$). A value of China (CHN, $446$) means that there are $446$ author entries from China in the dataset. The graph in Figure \ref{top20} displays the scholarly output of countries with a ranking of $11$ to $20$: Singapore (SGP, $33$), Finland (FIN, $31$), Austria (AUT, $29$), New Zealand (NZ, $25$), Canada (CAN, $17$), Switzerland (CHE, $16$), Italy (ITA, $13$), United Kingdom (UK, $13$), Iran (IRN, $12$) and Norway (NOR, $12$). We observe that Netherland, Malaysia, Morocco, Greece, Luxembourg and Turkey have a score of $8$, $7$, $7$, $5$, $5$ and $4$ respectively. There are several countries with a low scholarly output: Chile ($3$), Pakistan ($3$), Czech Republic ($2$), Portugal ($2$), Spain ($2$), Thailand ($2$), Vietnam ($2$), Denmark ($1$), Mexico ($1$) and Sudan ($1$). 
Table \ref{countrypoint} displays data on country of author and respective frequency over a period of $6$ years. Table \ref{countrypoint} reveals that China and Japan are the top two most contributing contributing countries to APSEC. Table \ref{countrypoint} shows that most of the countries are from Asia-Pacific region except Canada, Finland, France and USA. 
\begin{table*}[htpb]
		\centering
		\caption{Country of Author and Respective Frequency Data }
		\label{countrypoint}
				\renewcommand{\arraystretch}{1.1}
				\resizebox{18.5cm}{!}{
		\begin{tabular}{|l|c|l|c|l|c|l|c|l|c|l|c|}
			\hline
			\multicolumn{2}{|c|}{\textbf{2010}}  & \multicolumn{2}{|c|}{\textbf{2011}} & \multicolumn{2}{|c|}{\textbf{2012}} &\multicolumn{2}{|c|}{\textbf{2013}} & \multicolumn{2}{|c|}{\textbf{2014}} & \multicolumn{2}{|c|}{\textbf{2015}}  \\ \hline
			 \textbf{Country} & \textbf{Frequency(\%)} & \textbf{Country} & \textbf{Frequency(\%)} & \textbf{Country} & \textbf{Frequency(\%)}
			& \textbf{Country} & \textbf{Frequency(\%)} & \textbf{Country} & \textbf{Frequency(\%)}  & \textbf{Country} & \textbf{Frequency(\%)}    \\ \hline
		China & 56 (33.94) & China & 39 (23.78) & China & 123 (29.85) & Japan & 68 (23.69) & China & 100 (39.22) & China & 70 (41.67) \\ \hline
		Japan & 25 (15.15) & Japan & 36 (21.95) & Japan & 77 (18.69) & China & 58 (20.21) & Japan & 52 (20.39) & India & 29 (17.26) \\ \hline
		Australia & 20 (12.12) & France & 13 (7.93) & India & 29 (7.04) & India & 27 (9.41) & South Korea & 33 (12.94) & Japan & 24 (14.29) \\ \hline
		USA & 14 (8.48) & Australia & 12 (7.32) & France & 26 (6.31) & France & 20 (6.97) & India & 17 (6.67) & Australia & 9 (5.36) \\ \hline
		New Zealand & 10 (6.06) & Canada & 8 (4.88) & Taiwan & 20 (4.85) & Finland & 17 (5.92) & France & 11 (4.31) & South Korea & 7 (4.17) \\ \hline
		\end{tabular}}
	\end{table*}
\section{Prolific Author, Author Churn, Average Author Per paper}
\begin{table*}[htpb]
	\centering
	\caption{List of Most Prolific Authors (Country of Affiliation, Number of Papers Published and Total Number of Citations Received by All Published Papers) from APSEC $2010$ to $2015$}
	\label{prolificauthor}
			\renewcommand{\arraystretch}{1.1}
			\resizebox{18.5cm}{!}{
	\begin{tabular}{|l|l|c|c|c|c|l|l|c|c|c|c|}
		\hline
		\textbf{Name} & \textbf{Country} & \textbf{\# PR} & \textbf{\# CT} & \textbf{\# YR} & \textbf{\# CL} & \textbf{Name} & \textbf{Country} & \textbf{\# PR} & \textbf{\# CT} & \textbf{\# YR} & \textbf{\# CL}   \\ \hline			
Mikio Aoyama & Japan & 11 & 22 & 1,1,1,3,3,2 & 27 & Xiaohong Chen & China & 5 & 7 & 0,1,0,1,1,2 & 17  \\ \hline
Ashish Sureka & India & 8 & 70 & 1,1,1,1,3,1 & 15 & Shinpei Hayashi & Japan & 5 & 19 & 1,1,1,2,0,0 & 16  \\ \hline
Jing Liu & China & 7 & 9 & 0,1,1,1,2,2 & 21 & Li Zhang & China & 5 & 9 & 0,1,1,1,1,1 & 10  \\ \hline
Xiaoxing Ma & China & 6 & 20 & 0,0,2,1,2,1 & 28 & Kazuhiro Ogata & Japan & 5 & 2 & 0,0,2,1,1,1 & 7  \\ \hline
Shaoying Liu & Japan & 6 & 21 & 0,1,2,0,1,2 & 5 & Jian Lu & China & 5 & 22 & 0,0,2,0,3,0 & 27  \\ \hline
Motoshi Saeki & Japan & 6 & 30 & 2,1,1,2,0,0 & 20 & Horst Lichter & Germany & 5 & 20 & 0,0,2,1,2,0 & 9  \\ \hline
Chun Cao & China & 6 & 18 & 0,0,2,1,1,2 & 25 & Hironori Washizaki & Japan & 5 & 8 & 0,0,0,1,3,1 & 35  \\ \hline
Chang Xu & China & 6 & 22 & 0,0,2,0,3,1 & 31 & Hee Beng Kuan Tan & Singapore & 5 & 4 & 0,1,2,2,0,0 & 9  \\ \hline
Bixin Li & China & 6 & 25 & 2,0,4,0,0,0 & 22 & Ewan Tempero & New Zealand & 5 & 185 & 1,1,1,1,0,1 & 15  \\ \hline
Zhi Jin & China & 5 & 24 & 1,1,2,1,0,0 & 15 & Atul Gupta & India & 5 & 22 & 0,0,2,3,0,0 & 9  \\ \hline
Yoshiaki Fukazawa & Japan & 5 & 8 & 0,0,0,1,3,1 & 35 &  &  &  &  &  &   \\ \hline
\end{tabular}}
\end{table*}
\begin{table}[t]
\centering
\caption{Total Number of Unique Authors (UNAUTH), Number (NUMNEW) and Percentage of New Authors Added Every Year (PERNEW)}
\label{uniqueauthor}
\begin{tabular}{|l|*{4}{c|}r}
\hline
\textbf{Year} & \textbf{UNAUTH} & \textbf{NUMNEW} & \textbf{PERNEW}  \\ \hline
2010 & 157 & 0 &  \\ \hline
2011 & 158 & 147 & (93.04 \%) \\ \hline
2012 & 359 & 316 & (88.02 \%) \\ \hline
2013 & 271 & 184 & (67.90 \%) \\ \hline
2014 & 220 & 168 & (76.36 \%) \\ \hline
2015 & 157 & 119 & (75.80 \%) \\ \hline
\end{tabular}
\end{table}
\begin{table}[ht]
\caption{Average Authors Per Paper}
\label{avgauthor}
\centering
\begin{tabular}{|c|c|c|c|c|} 
\hline
\textbf{Year} & \textbf{Min.} & \textbf{Max.} & \textbf{Mean} & \textbf{Median}  \\ \hline
2010 & 1 & 8 & 3.51 & 3 \\ \hline
2011 & 1 & 6 & 3.35 & 3 \\ \hline
2012 & 1 & 7 & 3.38 & 3 \\ \hline
2013 & 1 & 7 & 3.30 & 3 \\ \hline
2014 & 1 & 13 & 3.70 & 3 \\ \hline
2015 & 1 & 8 & 3.29 & 3 \\ \hline
\end{tabular}
\end{table}
%
%
%
%
%
%
%
%
Table \ref{prolificauthor} shows the list of most prolific authors (from $2010$ to $2015$), the number of papers published by them (\#PR) and the total number of citations (\#CT) received by the published papers. Table \ref{prolificauthor} also displays the count of papers published in each year (\#YR). For example, Ashish Sureka published $1$, $1$, $1$, $1$, $3$ and $1$ paper(s) in the year $2010$, $2011$, $2012$, $2013$, $2014$ and $2015$ respectively. We compute the total number of collaborating co-authors for each of the prolific authors listed in Table \ref{prolificauthor}. 

We observe form Table \ref{prolificauthor} that higher productivity does not mean higher impact as some authors have a less papers in comparison to others but a relatively much higher impact. Table \ref{uniqueauthor} displays the number of unique authors and number of new authors added every year with respect to all the previous years in the dataset. We calculate the author churn and observe that the author churn is more than $65\%$ every year which means than at-least more than $65\%$ of the authors are new in comparison to previous years. Table \ref{avgauthor} shows the minimum, maximum, mean and median number of authors every year. Table \ref{avgauthor} reveals that the median value for the number of authors per paper is $3$ for the $6$ years. Our analysis reveals that there are several solo-authored papers as well as papers involving large collaborations (more than $6$ co-authors per paper). 
\begin{table}[h]
\centering
\caption{Number and Percentage of Research Track Papers having All Authors from University (AU), All Authors from Industry (AI) and Authors from both University and Industry (UI)}
\label{uicollab}
\begin{tabular}{|l|*{5}{c|}r}
\hline
\textbf{Year} & \textbf{NUM} & \textbf{AU} & \textbf{AI} & \textbf{UI} \\ \hline
2010 & 47 & 42  (89.36 \%)& 1  (2.13 \%)& 4  (8.51 \%)\\ \hline
2011 & 49 & 38 (77.55 \%)& 0  (0.00 \%)& 11  (22.45 \%)\\ \hline
2012 & 122 & 96  (78.69 \%)& 11  (9.02 \%)& 15  (12.30 \%)\\ \hline
2013 & 87 & 64  (73.56 \%)& 5  (5.75 \%)& 18  (20.69 \%)\\ \hline
2014 & 69 & 54  (78.26 \%)& 7  (10.14 \%)& 8  (11.59 \%)\\ \hline
2015 & 51 & 39  (76.47 \%)& 4  (7.84 \%)& 8  (15.69 \%)\\ \hline
\end{tabular}
\end{table}		
\section{Collaboration (University-Industry, Internal-External)}
Joint authorship in scientific papers is an evidence of collaboration and interaction between researchers as well as institutions. Our objective is to study university-industry collaboration and knowledge flow between the two types of institutions. Table \ref{uicollab} displays the data on university-industry collaboration. Table \ref{uicollab} reveals that the percentage of joint university-industry papers varies from a minimum of $8.51\%$ to a maximum of $22.45\%$. 

\begin{figure}[th]
\centering
\includegraphics[width=0.5\textwidth]{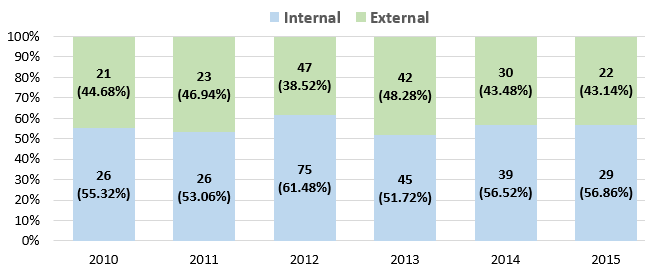}
\caption{Stacked Bar Chart Indicating the Percentage Distribution of Internal and External Collaboration}
\label{internalexternal}
\end{figure}

We investigate the nature and scale of collaboration in APSEC papers (indicated by joint authorship) from the perspective of internal or external collaboration. Internal collaboration is a form of collaboration in which all the co-authors in a paper (single or multiple-authors) are affiliated to one Institution only. External collaboration is defined as a form of collaboration which involves participation of two or more institutions (irrespective of whether the organizations involved are industry or university) in the production of the scientific output and the paper. Figure \ref{internalexternal} displays a stacked bar chart indicating the percentage distribution of internal and external collaboration. Figure \ref{internalexternal} reveals a good (above $35\%$) percentage of external collaboration. The percentage of papers having external collaboration varies from a minimum of $38.52\%$ to a maximum of $46.94\%$. 
\section{Program Committee Characteristics}
%
%
%
\begin{table}[h]
	\centering
	\caption{Program Committee Characteristics: Number, Diversity in-terms of Countries and Churn}
	\label{programcountry}
		\begin{tabular}{|l|*{3}{c|}r}
			\hline
			\textbf{Year} & \textbf{NUM} & \textbf{ Country} & \textbf{NEW} \\ \hline
			2010 & 83 & 18 (21.69 \%) & -\\ \hline
			2011 & 95 & 18 (18.95 \%) & 40 (48.19 \%)\\ \hline
			2012 & 103 & 18 (17.48 \%)& 54 (56.84 \%)\\ \hline
			2013 & 83 & 17 (20.48 \%)& 63 (61.16 \%)\\ \hline
			2014 & 65 & 19 (29.23 \%)& 30 (46.15 \%)\\ \hline 
			2015 & 75 & 14 (18.67 \%)& 8 (10.66 \%)\\ \hline
		\end{tabular}
	\end{table}			

The technical program committee for any conference is responsible for reviewing submitted papers and ensuring good quality reviews as well as selection of high quality papers. The size of the program committee should be according to the number of papers normally received by the conference so that the workload of the program committee members is reasonable or moderate. We extract the size of the program committee (for both the research and industry track) from the APSEC conference proceedings. Table \ref{programcountry} shows that the number of program committee members varies from a minimum of $65$ in year $2014$ to a maximum of $103$ in $2012$. The number of papers submitted at APSEC is in the range of $160$ to $230$ and hence the distribution of workload is moderate. APSEC is a long-running international conferences attracting papers from different parts of the world both from industry and academia. Diversity of institution, technical area of expertise and country is an important selection criteria for selecting program committee member and is an indicator of the quality of a conference. We extract the country of every program committee member and compute the number of different countries. Table \ref{programcountry} shows that APSEC program committee is diverse and inclusive in-terms of the number of countries. For example, in the year $2010$, there were $83$ program committee members from $18$ different countries (a diversity score of $21.69\%$). In year $2014$, there were $65$ members from $19$ countries. The study by Vasilescu et al. reveals that the lowest observed PC renewal ratio is $8.8\%$ for SCAM 2005, and the highest is $93\%$ for GPCE 2007 \cite{vasilescu2014}. Vasilescu et al. analysis shows that wider-scope software engineering conferences such as ICSE, FSE and FASE have a relatively higher PC turnover than narrow-scoped conferences \cite{vasilescu2014}. Their analysis does not include APSEC. APSEC falls in the category of a wider scope software engineering conferences. 

\begin{figure}[th]
\centering
\includegraphics[width=0.47\textwidth]{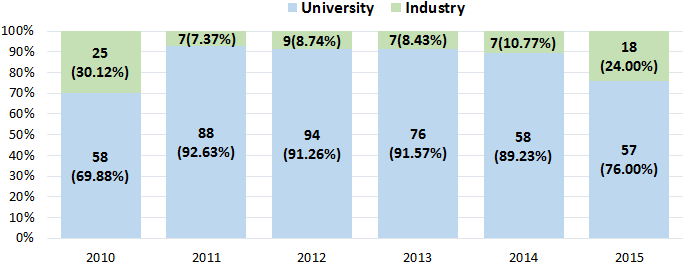}
\caption{Percentage Distribution of Program Committee Members across Industry and University}
\label{pcunivind}
\end{figure}

Annual churn and rotation of program committee members is essential for making sure that there is diversity, inclusiveness and cross-section of topic expertise, institution and geographical area. Inviting new program committee members and making space for them by rotating-off program committee members who have served for $2-3$ years are normal guidelines for conferences. We compute the yearly churn in the program committee for APSEC $2010$ to APSEC $2015$. Table \ref{programcountry} shows that in the year $2011$ there were a total of $95$ program committee members out of which $40$ ($48.19\%$) were new and $55$ were repeated from the year $2010$. We observe that the highest churn was in the year $2013$ ($61.16\%$) and the lowest was in the year $2015$ ($10.66\%$). One of the objectives of conferences like APSEC is to stimulate interaction among researchers in industry and academia and provide a forum for computing professionals from both industry and academia to present and exchange research results. Hence, it is an important guideline for program committee chairs who lead the program committee member selection and invitation process to have a balanced representation from both industry and academia. We extract the affiliation of each program committee member and determine whether the member belongs to an industry or university.  Table \ref{pcunivind} displays (as a stacked bar-chart) the percentage distribution of program committee members across industry and university. Table \ref{pcunivind} reveals an imbalance between industry and academia and is skewed towards university. The percentage of program committee members from university varies from a minimum of $69.88\%$ to a maximum of $92.63\%$. We observe that for $3$ years ($2010$, $2011$ and $2012$), the percentage of program committee members from industry is less than $10\%$. We notice an upward trend in percentage of program committee members from industry for past $3$ years ($8.43\%$ to $24\%$). 
%

%

\section{Gender Imbalance in Authorship}
\begin{figure}[th]
\centering
\includegraphics[width=0.47\textwidth]{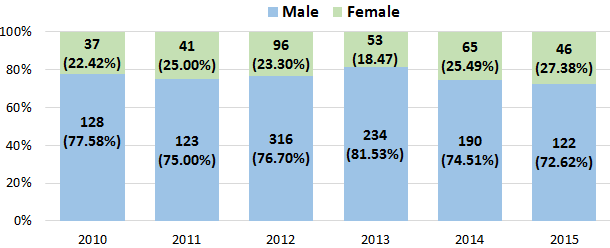}
\caption{Percentage of Male and Female Authors (Gender Imbalance)}
\label{gender}
\end{figure}

Agarwal et al. conduct an analysis of women in computer science research by analyzing author data from $81$ conferences including $11$ conferences in software engineering \cite{swati2016}. Their experimental dataset consists of DBLP bibliography entries from the year $2000$ to $2015$. Their results reveal that $79\%$ of the authors in the bibliography dataset consisting of $11$ conferences and $16$ years are male whereas $21\%$ authors are women authors \cite{swati2016}. We use the Genderize.io\footnote{\url{https://genderize.io/}} API to determine the gender of all the authors in our dataset. The Genderize.io database (as of $10$ June $2016$) contains $216286$ distinct names across $79$ countries and $89$ languages and can be used to determine the gender of a first name. The Genderize.io API returns the gender with a value of null if it is not able to determine the gender from a given first name. We manually checked the gender for $405$ author names by doing a web search and checking the author profiles such as their homages consisting of pictures. Figure \ref{gender} shows the percentage of male and female authors for APSEC $2010$ to APSEC $2015$.  Figure \ref{gender} reveals a gender imbalance as the percentage of women authors varies from a minimum of $18.47\%$ (year $2013$) to a maximum of $27.38\%$ (year $2015$). We observe an increasing trend over past $3$ years however there has been decline in percentage of women authors from year $2011$ to $2012$ and $2012$ to $2013$. We also observe that there has never been a women General Chair out of the $11$ General Chairs in APSEC from the year $2010$ and $2015$.  
\section{Topic Analysis and Distribution}
Our objective is to identify major research topics and trends in APSEC $2010$ to APSEC $2015$. We extract all the keywords from the BibTeX citation entries of every paper in our dataset. The keywords contains terms from Inspec (a major indexing database of scientific and technical literature) controlled and non-controlled indexing, author keywords and IEEE terms. \footnote{\url{ https://www.computer.org/web/swebok}}. We map each key-term to one of the $15$ knowledge areas (KAs) listed in the Guide to the Software Engineering Body of Knowledge (SWEBOK Guide). The Guide to the Software Engineering Body of Knowledge (SWEBOK Guide) describes generally accepted knowledge about software engineering. We use SWEBOK V3.0 which is the most recent completely revised and updated version as of $11$ June $2016$. The $15$ KAs are broad topics such as software requirements, software design, software construction, software testing and software maintenance. The SWEBOK V3.0 defines sections and sub-sections (sub-topics) within each KA. Table \ref{topicanalysis} displays the $15$ KAs listed in SWEBOK Guide and the mapping of key-terms present in the BibTeX citation entry of every paper to the KAs. Table \ref{topicanalysis} reveals the top two knowledge areas for every year marked in bold and with a superscript. Our analysis reveals that software design is the most popular knowledge area. The second most popular knowledge area for the year $2011$ and $2012$ is software testing and software construction respectively. Overall (cumulative sum of all the $6$ years) the top five knowledge areas are:  software design, computing foundations, software testing, software construction and mathematical foundations. Our analysis reveals that there are not many papers in the area of engineering foundations, software engineering professional practice and software engineering models and methods. Mathew et al. conduct a topic modeling analysis of the abstracts and titles from $9291$ papers published in $11$ top-ranked SE conferences between $1993$ to $2013$ \cite{mathew2016}. Their analysis does not include APSEC. Their analysis reveals that software testing is the most dominant theme in modern software engineering. Our results on APSEC topic modeling shows that software testing is also one of the most dominant area. Garousi et al. classify the top software engineering papers by their coverage of SE topics \cite{garousi2016highly}. They adopted the classification of SE areas presented in the 2014 version of the SWEBOK. Their analysis revels that models and methods and software design knowledge areas are the most popular topics in the top papers (in 50 and 34 of the top 135 papers, respectively) \cite{garousi2016highly}.
\begin{table}[htpb]
\centering
\caption{$15$ Knowledge Areas and the Mapping of Key-Terms Present in the BibTeX Citation Entry}
\label{topicanalysis}
\renewcommand{\arraystretch}{1.1}
\resizebox{8.5cm}{!}{
\begin{tabular}{|p{4.0cm}|c|c|c|c|c|c|c|}
\hline
\textbf{Topic} & \textbf{2010} & \textbf{2011} & \textbf{2012} & \textbf{2013} & \textbf{2014} & \textbf{2015} & \textbf{Sum} \\ \hline
Software Requirements & 87 & 88 & 208 & 146 & 123 & 37 & 647  \\ \hline
Software Design & \textbf{$1391^{1}$} & \textbf{$177^{1}$} & \textbf{$438^{1}$} & \textbf{$319^{1}$} & \textbf{$276^{1}$} & \textbf{$133^{1}$} & \textbf{$133^{1}$}  \\ \hline
Software Construction & 75 & 100 & \textbf{$334^{2}$} & 214 & 132 & 53 & 936  \\ \hline
Software Testing & 95 & \textbf{$133^{2}$} & 286 & 206 & 151 & 59 & 961  \\ \hline
Software Maintenance & 90 & 117 & 326 & 202 & 157 & 44 & 795  \\ \hline
Software Configuration Management & 43 & 38 & 116 & 85 & 60 & 12 & 354  \\ \hline
Software Engineering Management & 41 & 48 & 87 & 53 & 48 & 10 & 287  \\ \hline
Software Engineering Process & 76 & \textbf{$92^{2}$} & 184 & 134 & 115 & 46 & 908  \\ \hline
Software Engineering Models and Methods & 10 & 26 & 61 & 36 & 42 & 5 & 155  \\ \hline
Software Quality & 71 & 96 & 270 & 170 & 150 & 38 & 689  \\ \hline
Software Engineering Professional Practice & 9 & 13 & 32 & 10 & 14 & 5 & 83  \\ \hline
Software Engineering Economics & 13 & 21 & 44 & 33 & 35 & 9 & 180  \\ \hline
Computing Foundations & 93 & 97 & 310 & 219 & \textbf{$181^{2}$} & \textbf{$61^{2}$} & \textbf{$1007^{2}$}  \\ \hline
Mathematical Foundations & \textbf{$113^{2}$} & 107 & 329 & \textbf{$232^{2}$} & 174 & 52 & 930  \\ \hline
Engineering Foundations & 0 & 0 & 0 & 0 & 0 & 0 & 0  \\ \hline
\end{tabular}}
\end{table}

We categorize every paper in our dataset into one of the three categories: AU (All University), AI (All Industry) and UI (University Industry). AU consists of papers in which all the co-authors are affiliated to university. UI consists of papers in which co-authors are affiliated to university as well as industry. Table \ref{top3topics} displays top $3$ topics for all university authored papers, all industry authored papers and university and industry collaboration papers. The information in Table \ref{top3topics} provides insights on topics which are investigated more in industry in comparison to the topics in university and vice-a-versa. Similarly, our objective is to examine the research topics which are more common in papers involving university and industry collaboration. Our analysis reveals that software design is the most popular knowledge area in all the $3$ categories. We observe that software testing is the second most popular knowledge area for the UI category and is not present in the top $3$ list in AU category except for the year $2011$. We notice that software engineering process is not present in the top $3$ list within the AU category but is present two times in the AI category (year $2014$ and $2015$). 
\begin{table}[htpb]
\centering
\caption{Top $3$ Topics for All University Authored Papers, All Industry Authored Papers and University and Industry Collaboration Papers}
\label{top3topics}
\renewcommand{\arraystretch}{1.1}
\resizebox{8.5cm}{!}{
\begin{tabular}{|c|l|l|l|}
\hline
\textbf{Year} & \textbf{1st} & \textbf{2nd} & \textbf{3rd}  \\ \hline
\multicolumn{4}{|c|}{\textbf{All University}} \\ \hline
2010 & Software Design & Mathematical Foundations & Software Testing  \\ \hline
2011 & Software Design & Software Testing & Software Maintenance  \\ \hline
2012 & Software Design & Software Maintenance & Software Construction  \\ \hline
2013 & Software Design & Mathematical Foundations & Software Maintenance  \\ \hline
2014 & Software Design & Computing Foundations & Mathematical Foundations  \\ \hline
2015 & Software Design & Computing Foundations & Software Construction  \\ \hline
All & Software Design & Mathematical Foundations & Computing Foundations  \\ \hline
\multicolumn{4}{|c|}{\textbf{All Industry}} \\ \hline

2010 & Mathematical Foundations & Software Requirements & Software Maintenance  \\ \hline
2011 & No Papers & No Papers & No Papers  \\ \hline
2012 & Software Design & Software Quality & Computing Foundations  \\ \hline
2013 & Software Testing & Mathematical Foundations & Software Requirements  \\ \hline
2014 & Software Design & Software Maintenance & Software Engineering Process  \\ \hline
2015 & Software Quality & Computing Foundations & Software Engineering Process  \\ \hline
All & Software Design & Software Testing & Computing Foundations  \\ \hline
\multicolumn{4}{|c|}{\textbf{Both (University and Industry)}} \\ \hline

2010 & Software Design & Software Engineering Process & Computing Foundations  \\ \hline
2011 & Software Design & Mathematical Foundations & Software Testing  \\ \hline
2012 & Software Design & Software Construction & Mathematical Foundations  \\ \hline
2013 & Software Design & Software Testing & Software Construction  \\ \hline
2014 & Software Design & Computing Foundations & Software Quality  \\ \hline
2015 & Software Design & Software Testing & Software Engineering Process  \\ \hline
All & Software Design & Software Testing & Mathematical Foundations  \\ \hline
\end{tabular}}
\end{table}
\\\\
\textbf{APSEC 2016:} received 218 submissions (originally 225 with 7 withdrawn) of which the program committee accepted 43 papers as regular papers and 20 papers as short papers giving a $28.9\%$ acceptance rate for both regular and short papers. The short version of this paper (A Review of Six Years of Asia-Pacific Software Engineering Conference) is accepted in APSEC 2016 happening in University of Waikato, Hamilton, New Zealand. The conference accepted three workshop proposals: International Workshop on Quantitative Approaches to Software Quality, Case Method for Computing Education (CMCE): A Strategy for Teaching Software Engineering and First International Workshop on Technical Debt Analytics. The keynote speakers and their topics are: (1) Cristina Cifuentes (Research Director, Oracle Labs Australia), Oracle Parfait: The Flavour of Real-World Vulnerability Detection, (2) Manu Sridharan (Samsung Research America), Program Analysis for Real-World JavaScript, (3) Paul Ash (Director, National Cyber Policy Office, Department of Prime Minister and Cabinet, New Zealand Government), New Zealand in an interconnected world: delivering a secure, resilient and prosperous online environment. The conference accepted six tutorials: (1) Weka Data Mining Tool (2) Applying Model Driven Engineering Technologies in the Creation of Domain Specific Modeling Languages (3) SOUFFLÉ: Datalog Compiler for Static Program Analysis (4) Interaction Design for Specifying Requirements (5) Learn to Build and Apply Software Analysis Tools with Interactive Visualization (6) Dynamic analysis of JavaScript with Jalangi. The conference is sponsored by: Oracle Labs, Google, Massey University, Rezare Systems, The University of Waikato, AUT Software Engineering Research Laboratory and the University of Auckland. The conference General Chairs are (1) Jens Dietrich, Massey University, New Zealand (2) Steve Reeves, University of Waikato, New Zealand. The Research Program Track Chairs are (1) Gail Murphy, University of British Columbia, Canada (2) Alex Potanin, Victoria University of Wellingon, New Zealand. The program committee consists of $51$ members.
\section{Conclusions}
We conclude that APSEC  is successfully meeting its desired objective as it is able to attract a good number of papers ($150$ - $230$) from different parts of the world both from industry and academia. The acceptance rate (between $30\%$ - $35\%$ on an average) demonstrates that APSEC is a moderately selective conference. APSEC has a good quality co-located workshops and tutorials covering diverse topics. The citation impact of the conference is moderate indicating that APSEC is maintaining its status as a Tier $2$ conference. The papers published in APSEC demonstrates both university-industry collaboration as well as external collaboration. The scholarly output from various countries at APSEC shows that APSEC is able to attract good number of submissions from authors form different geographical regions in the world. The program committee of APSEC is diverse both from the perspective of representations from industry and academia and from different countries. There is a healthy program committee and author churn which indicates that the conference is broad and open.  
\bibliographystyle{IEEEtran}  
\bibliography{apsec}  

\begin{thebibliography}{10}
\providecommand{\url}[1]{#1}
\csname url@samestyle\endcsname
\providecommand{\newblock}{\relax}
\providecommand{\bibinfo}[2]{#2}
\providecommand{\BIBentrySTDinterwordspacing}{\spaceskip=0pt\relax}
\providecommand{\BIBentryALTinterwordstretchfactor}{4}
\providecommand{\BIBentryALTinterwordspacing}{\spaceskip=\fontdimen2\font plus
\BIBentryALTinterwordstretchfactor\fontdimen3\font minus
  \fontdimen4\font\relax}
\providecommand{\BIBforeignlanguage}[2]{{%
\expandafter\ifx\csname l@#1\endcsname\relax
\typeout{** WARNING: IEEEtran.bst: No hyphenation pattern has been}%
\typeout{** loaded for the language `#1'. Using the pattern for}%
\typeout{** the default language instead.}%
\else
\language=\csname l@#1\endcsname
\fi
#2}}
\providecommand{\BIBdecl}{\relax}
\BIBdecl

\bibitem{tripathi2015}
A.~Tripathi, S.~Dabral, and A.~Sureka, ``University-industry collaboration and
  open source software (oss) dataset in mining software repositories (msr)
  research,'' in \emph{Software Analytics (SWAN), 2015 IEEE 1st International
  Workshop on}, 2015, pp. 39--40.

\bibitem{sharma2016}
R.~Sharma, P.~Aggarwal, and A.~Sureka, ``Insights from mining eleven years of
  scholarly paper publications in requirements engineering (re) series of
  conferences,'' \emph{SIGSOFT Softw. Eng. Notes}, vol.~41, no.~2, pp. 1--6,
  May 2016.

\bibitem{swati2016}
S.~Agarwal, N.~Mittal, R.~Katyal, A.~Sureka, and D.~Correa, ``Women in computer
  science research: What is the bibliography data telling us?'' \emph{SIGCAS
  Comput. Soc.}, vol.~46, no.~1, pp. 7--19, Mar. 2016.

\bibitem{barn2016}
B.~S. Barn, T.~Clark, A.~Ali, and R.~Arif, ``A systematic mapping study of the
  current practice of indian software engineering,'' in \emph{Proceedings of
  the 9th India Software Engineering Conference}, ser. ISEC '16, 2016, pp.
  89--98.

\bibitem{freitas2011}
F.~G. Freitas and J.~T. Souza, \emph{Search Based Software Engineering: Third
  International Symposium, SSBSE 2011, Szeged, Hungary, September 10-12, 2011.
  Proceedings}, 2011, ch. Ten Years of Search Based Software Engineering: A
  Bibliometric Analysis, pp. 18--32.

\bibitem{bergamaschi2012}
R.~A. Bergamaschi, R.~C. Rezende, H.~P. De~Oliveira, and A.~Kumon, Jr., ``A
  quantitative analysis of www, hypertext and jcdl conferences in the last
  decade,'' \emph{SIGWEB Newsl.}, no. Winter, pp. 5:1--5:8, Jan. 2012.

\bibitem{bartneck2010}
C.~Bartneck, ``The end of the beginning: a reflection on the first five years
  of the hri conference,'' \emph{Scientometrics}, vol.~86, no.~2, pp. 487--504,
  2010.

\bibitem{bartneck2009}
C.~Bartneck and J.~Hu, ``Scientometric analysis of the chi proceedings,'' in
  \emph{Proceedings of the SIGCHI Conference on Human Factors in Computing
  Systems}, ser. CHI '09, 2009, pp. 699--708.

\bibitem{agarwalsigweb2016}
\BIBentryALTinterwordspacing
S.~Agarwal, N.~Mittal, and A.~Sureka, ``A glance at seven acm sigweb series of
  conferences,'' \emph{SIGWEB Newsl.}, no. Summer, pp. 5:1--5:10, Jul. 2016.
  [Online]. Available: \url{http://doi.acm.org/10.1145/2956573.2956578}
\BIBentrySTDinterwordspacing

\bibitem{sureka2016}
\BIBentryALTinterwordspacing
A.~Sureka, ``{Meta-Data of Papers in Asia-Pacific Software Engineering
  Conference from 2010 to 2015},'' 9 2016. [Online]. Available:
  \url{https://figshare.com/articles/Meta-Data_of_Papers_in_Asia-Pacific_Software_Engineering_Conference_from_2010_to_2015/3843996}
\BIBentrySTDinterwordspacing

\bibitem{lov2016}
L.~Kumar, S.~Sripada, and A.~Sureka, ``A review of six years of asia-pacific
  software engineering conference,'' in \emph{Asia-Pacific Software Engineering
  Conference (APSEC)}, 2016.

\bibitem{garousi2016highly}
V.~Garousi and J.~M. Fernandes, ``Highly-cited papers in software engineering:
  The top-100,'' \emph{Information and Software Technology}, vol.~71, pp.
  108--128, 2016.

\bibitem{vasilescu2014}
B.~Vasilescu, A.~Serebrenik, T.~Mens, M.~G. van~den Brand, and E.~Pek, ``How
  healthy are software engineering conferences?'' \emph{Science of Computer
  Programming}, vol.~89, pp. 251--272, 2014.

\bibitem{mathew2016}
G.~Mathew and A.~Agarwal, ``Trends in topics at se conferences (1993-2013),''
  \emph{arXiv preprint arXiv:1608.08100}, 2016.

\end{thebibliography}
\end{document}